\documentclass[12pt]{article}
\usepackage{graphicx}
\usepackage{a4}
\usepackage{float}
\usepackage{hyperref}
\hypersetup{
    colorlinks=true,
    linkcolor=blue,
    filecolor=magenta,      
    urlcolor=cyan,
    pdftitle={Sharelatex Example},
    bookmarks=true,
    pdfpagemode=FullScreen,
}
\begin{document}


\thispagestyle{empty}
\setlength{\parindent}{1.0cm}

\begin{center}
\bf 
Talk presented at the International Workshop on Future Linear Colliders 
(LCWS2016), Morioka, Japan, 5-9 December 2016. C16-12-05.4
\end{center}
\vskip 3cm
\begin{center}
\bf \Large CALICE Si/W ECAL: Endcap structures and cooling system
\end{center}

\vskip 3.cm
\begin{center}
{\Large
D. Grondin\footnotemark\footnotetext{Corresponding author. E-mail: Denis.Grondin@lpsc.in2p3.fr}, 
J. Giraud and J.-Y. Hostachy\\
{\it \large Laboratoire de Physique Subatomique et de Cosmologie - Universit\'{e}
Grenoble-Alpes, CNRS/IN2P3, Grenoble, France }\\}
\end{center}

\vskip 3.cm
\begin{center}
\bf \large Abstract
\end{center}
The next major project of particle physics will be the International 
Linear Collider: a linear accelerator in which electrons and  
positrons will collide with energies of 500 to around 1000  
billion  electronvolts.  
The LPSC-Grenoble is involved in the R\&D activities for the  
International Large Detector (ILD) and in particular in the electromagnetic 
calorimeter (ECal): i.e. design of the fastening and cooling  
systems, design of the mechanics of the end-caps, test of EM prototypes, 
tooling and integration of the end-caps.

\tableofcontents

\newpage
\section{Introduction}
The International Linear Collider (ILC) is a project for the future 
electron and positron collider which should be built in North of Japan.
Our contribution is essentially oriented towards research and development (R\&D)
in mechanics, in continuity with the history and the know-how
of electromagnetic (EM) calorimetry group in Grenoble, France.
We are members of the 
CALICE (CAlorimeter for the LInear Collider Experiment) 
collaboration~\cite{CALICE-Col}, 
which brings together 57 institutes,
17 different countries from four continents (Africa, America, Asia and Europe).

\section{R\&D in mechanics} 

The mechanical studies were carried out in collaboration with the 
LLR\footnotemark 
\footnotetext{Laboratoire Leprince-Ringuet} of Palaiseau, France
and the LAL\footnotemark 
\footnotetext{Laboratoire de l'Acc\'el\'erateur Lin\'eaire} d'Orsay, France.
These studies concern a silicon tungsten electromagnetic calorimeter featuring Carbon Fibre Reinforced Polymer
(CFRP)/Tungsten structures and silicon detectors, called briefly Si/W ECAL~\cite{R&D-ECAL}.

\subsection{General architecture of EM end-caps: advanced drawings and 
numerical simulations} 

\begin{figure}[H]
    \centering
    \includegraphics[width=15cm]{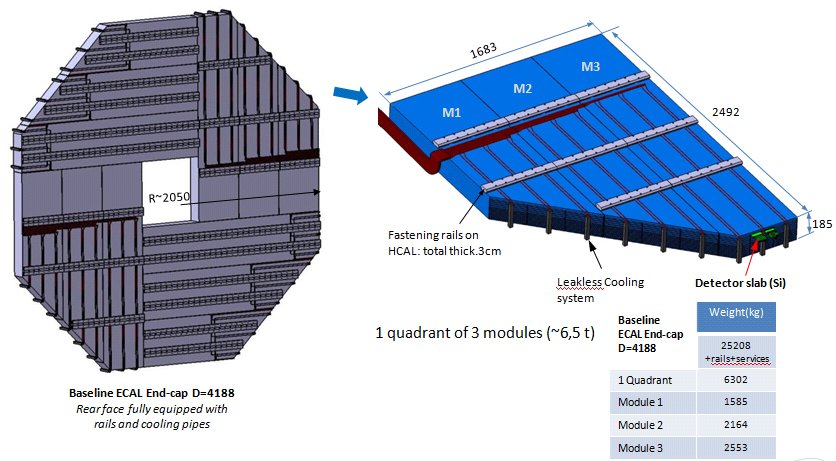}
    \caption{Design of the ECal end-cap.}
    \label{EndCap}
\end{figure}
\begin{figure}[h]
    \centering
    \includegraphics[width=8cm]{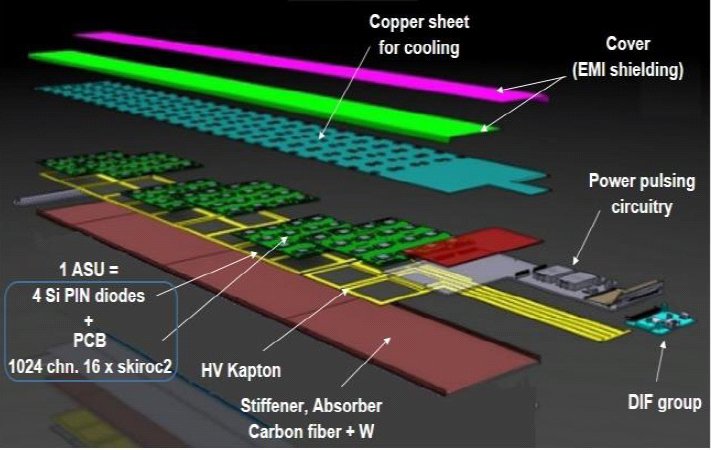}
    \caption{Exploded view of half a long slab.}
    \label{Slab}
\end{figure}

The weight of each end-cap is about 25.5 tons.
So numerical simulations have been undertaken to study the
mechanical behavior of this sub-detector divided 
in 4 quadrants composed of 3 modules each 
(see fig.~\ref{EndCap}). 
There are 3 alveoli rows per module, 
each containing 15 alveoli 
(i.e. 2$\times$540 alveoli in total).
The alveolar structure is made by moulding pre-impregnated carbon fibre
and epoxy onto tungsten sheets.
Free spaces of the alveolar structure allow to insert 
detector units called detector slabs (see fig.~\ref{Slab} and 
section~\ref{sec:LocCoolSys}).
One detector slab consists of two active layers (silicon diode wafers + PCB) 
mounted on each side of an H-shaped supportinng structure 
(including tungsten absorbers too),
and shielded on both sides with aluminum foils.
In addition, two copper sheets are placed on the two 
printed circuit boards (PCBs) of the active layers.
These copper sheets allow the heat transfer between electronic chips
embedded in the PCBs and the cooling heat exchanger located at the end 
of each slab.
 
\begin{figure}[h]
    \begin{minipage}[t]{7cm}
        \centering
        \includegraphics[width=6cm]{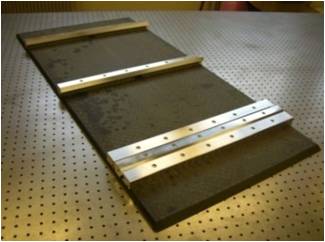}
        \caption{15 mm thick carbon HR plate with inserts 
	        and aluminum rails.}
	\label{ThickLayer}
    \end{minipage}
    \hspace{3mm}
    \begin{minipage}[t]{7cm}
        \centering
        \includegraphics[width=5cm]{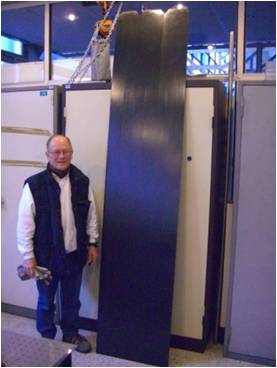}
        \caption{One layer molded (3 alveoli, L~=~2.490 m, 
	       wall thickness 0.5 mm).}
	\label{MouldingLayer}
    \end{minipage}
\end{figure} 

The mechanical stiffness of the composite support structure and 
the deformations
have already been the subject of studies followed by advanced drawings. 
The production of
components such as thick composite plates (13 mm thick) with metal inserts
for the fixing of the modules 
made it possible to
test the composite-metal interface system of the whole ECal on the hadronic 
calorimeter (see fig.~\ref{ThickLayer} and section~\ref{sec:Assembly}).

A new architecture that would avoid certain dead zones is being validated.
This new structure requires modules of about
2.5 m long and therefore requires the construction of prototypes in order to
test the feasibility of such a solution (see fig.~\ref{MouldingLayer}). 
The production of composite layers of three alveoli
revealed the need to evolve the molding.
Improvements on molds, cores, type and implementation of the prepreg 
have been carried out.

\begin{figure}[h]
    \centering
    \includegraphics[width=14cm]{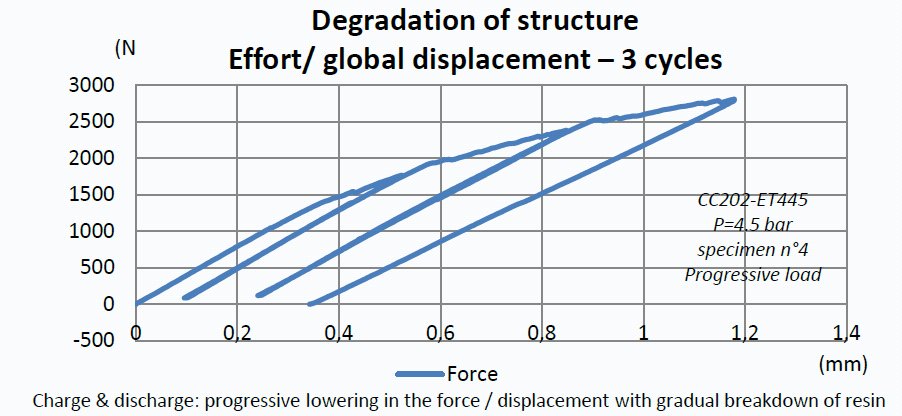}
    \caption{Shearing test: progressive loss of stiffness for the dummy structure.}
    \label{ShearingTest}
\end{figure}

Because of the problem of bending stress on the alveolar skin,  
a campaign of
shearing tests has been undertaken on representative specimens.
It showed a progressive loss of stiffness of the composite
structure (see fig.~\ref{ShearingTest}) which could occur 
due to the risk during integration and transport
and also seism.
The problem can be solved by increasing the number of external plies
but it will have an impact on the ECal dead zones.
The initial results validate in a first step the theoretical model of the
glued structures.
These tests will thus make it possible to adapt the simulation parameters
by finite elements in order to simulate the whole structure.
Additional tests of strain, on the resin in particular, will be conducted 
in order to validate definitively the model.

\subsection{Fastening of the electromagnetic calorimeter}
\label{sec:Assembly}

\begin{figure}[h]
    \begin{minipage}[t]{7cm}
        \centering
        \includegraphics[width=5.5cm]{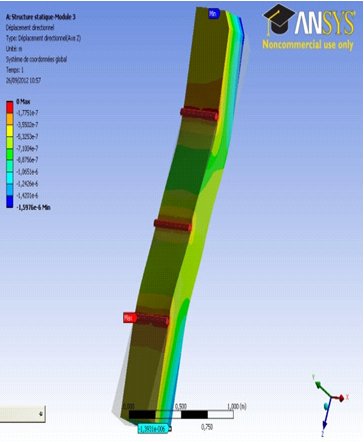}
        \caption{Composite simulation: optimisation of rail localisation 
                and module displacements.}
	\label{RailLocalisation}
    \end{minipage}
    \hspace{2mm}
    \begin{minipage}[t]{7cm}
        \centering
        \includegraphics[width=6cm]{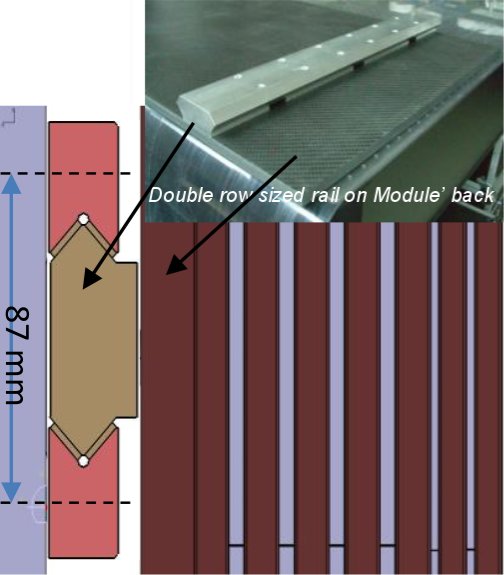}
        \caption{Fastening system on HCAL.}
        \label{Rail}
    \end{minipage}
\end{figure}

Only about 3 cm are available between the barrel or the end-caps of the 
electromagnetic and the hadronic calorimeter, for the passage of 
fluids and controls.
Therefore the fastening system is also a challenge.
Studies have been carried out to optimize the installation of the rails
(either in aluminum or composite material) 
in order to minimize the overall deformation of the various modules
(see fig.~\ref{RailLocalisation}).
This has led to the construction of double-row and small-section rails
(see fig.~\ref{Rail}), 
making it possible to secure the load on two rails for the majority 
of the 12 modules.

Also, the pull-out resistance of the rails and support inserts has been 
optimized in order to reduce the dead material at the rear of the modules.

\subsection{The local cooling system}
\label{sec:LocCoolSys}

Finally, a cooling device is necessary to evacuate the heat produced 
by the large number of electronic channels.
Here again the small space available (the electromagnetic calorimeter 
must remain as compact as possible)
is obviously a source of difficulty.
The cooling system is required to operate at temperatures between 
20$^{\circ}$C and 40$^{\circ}$C with a tolerance of $\pm$2.5$^{\circ}$C.
In order to meet
these requirements, a sub-atmospheric leak less water-cooling system 
has been chosen.
In case of leaks in the cooling
circuit, external air enters into the circuit preventing 
thus the cooling water to spill out.
Therefore, the detector electronics will not be exposed to water spray.

The current baseline design of ILD Si/W ECal~\cite{ILDBaseLine} 
with highly integrated layers designed for
minimal thickness and cooling capacity gives the global value 
of thermal power to be dissipated.
The system includes 40 identical barrel modules and 24 end-cap modules 
resulting in 8160 layers to be connected to heat exchangers. 
All slabs together contain $\sim$~75,000 Active Sensor Unit (ASU)
supporting 300,000 wafers and PCBs carrying 1.2 million SKIROC ASICs. 
The SKIROC ASIC dissipates
25~$\mu$W/channel for a total of 77 million channels.
Note at this point that the SKIROC ASICs will be
power pulsed with a duty cycle of 1\%. 

Taking into account power pulsing, the total heat dissipated 
by 256 ASICs and two DIF cards
amounts to $\sim$1~W per slab, 
it leads to around 4.6~kW for the whole Si/W ECAL. 
This heat must not
be released to the surrounding air nor being conducted to the detector itself, 
where the temperature has to be kept in 
the vicinity of 25$^{\circ}$C to 30$^{\circ}$C. 
Hence, one water circuit per column of each module
has been foreseen to remove the heat from the front end electronics. 

The copper drain is adapted to the Detector InterFace (DIF) card 
to be in contact with FPGA (i.e. Field-Programmable Gate Array).
With the nominal FPGA power of 300~mW, 
the power to be dissipated per slab in the barrel is 1~W.
The performance of the local cooling system has been simulated for both 
barrel and end-cap
configuration covering the power range of the FPGA (i.e. from 300~mW up to 2~W)
and meet thermal requirements.

\begin{figure}[H]
    \centering
    \includegraphics[width=15cm]{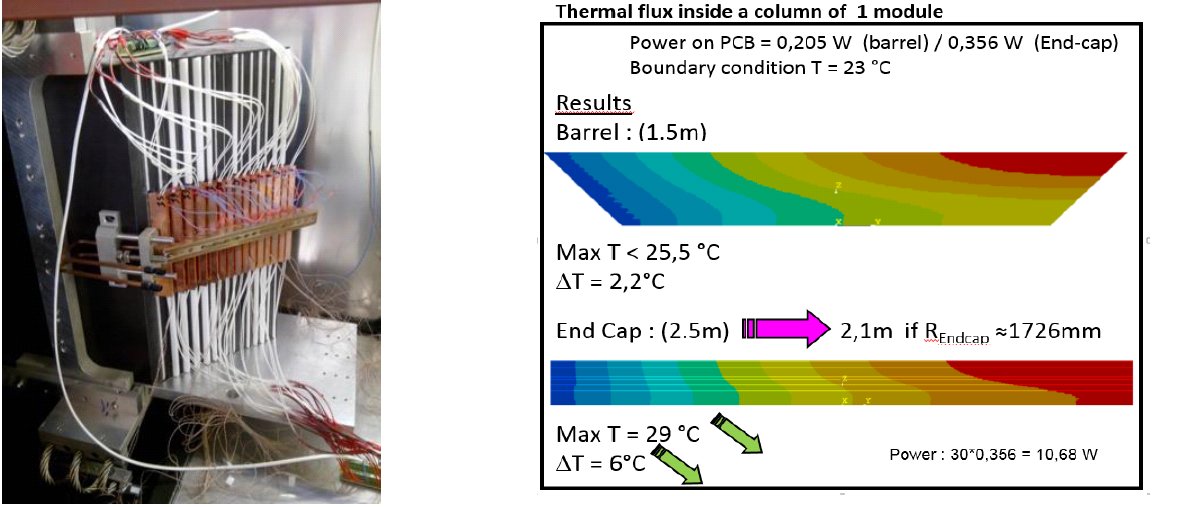}
    \caption{Left: EUDET module equipped for thermal tests with 15 dummy 
    slabs directly connected to the heat exchanger. Right: Thermal 
    analysis of module: the passive cooling is satisfactory with maximum 
    variation of temperature of 2.2$^{\circ}$C for the barrel modules 
    and 6$^{\circ}$C for the longest end-cap modules.}
    \label{ThermalTests}
\end{figure}

For the thermal tests, the so-called EUDET module was used 
(see fig.~\ref{ThermalTests}).
This technological prototype for a Si/W ECAL module 
(funded in part by the EUDET project)
corresponds in size to an ILD Si/W ECAL barrel module except 
for the number of columns (3 instead of 5).
The central column of the EUDET module is equipped with 15 slabs, 
which mimic the final local cooling system.
Thermal power values applied on slabs from 15~W to 30~W 
show the limit of heat load that can be produced
by evolution of electronics consumption.

The power dissipation is distributed along a detector slab 
concentrate at the location of the readout
ASICs. 
The goal is to transfer the residual heat as uniformly as possible 
from the ASU chain ($\sim$1500~mm length, 180 mm wide) up to a cold 
block located at one slab end. 
The challenge is the thermal
drain thickness, which should be close to the thickness 
obtained by simulations (i.e. 0.5~mm). 
The second aspect is to optimize the thermal contact between the copper 
and each ASIC, even if the
ASIC and its bonded wires are protected inside the PCB by an isolating resin. 
One proposed
solution is to use a thermal conductive agent.

Considering the power dissipation of the DIF board, 
a fastening assembly to the copper drain has been developed. 
To improve the heat transfer along the copper drain, 
in particular the local contact
between ASICs and the copper shield, 
several thermal conductive agents have been tested with
specialized companies. 
The most appropriate solution for now is conductive grease.

\begin{figure}[h]
    \centering
    \includegraphics[width=14cm]{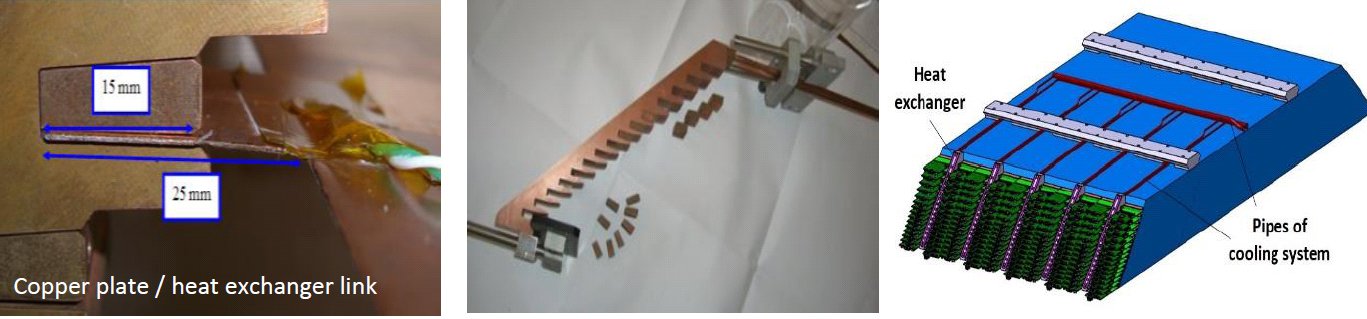}
    \caption{Left: Design of connection of heat exchangers with one 
    copper drain extremity. Middle: heat exchanger for 15$\times$2 connections. 
    Right: location of cooling system on one Si/W ECal barrel module.}
    \label{HeatExchanger2}
\end{figure}

For the connection of pipes to each slab, 
it is done by the way of cold copper blocs, 
brazed on pipes, inserted between the 2 copper sheets of the slab, 
in the free space let between the 2 DIF cards (see fig.~\ref{HeatExchanger2}). 
Then, there is one cooling water exchanger in front of each column.
The prototype has been dimensioned to fit EUDET module dimensions, 
with its specific shape (trapezoidal), and has been tested. 
The slabs are therefore cooled down only at the extremity which 
is the part that supports the most important number of electronic 
components to cool. 

\begin{figure}[!h]
    \centering
    \includegraphics[width=12cm]{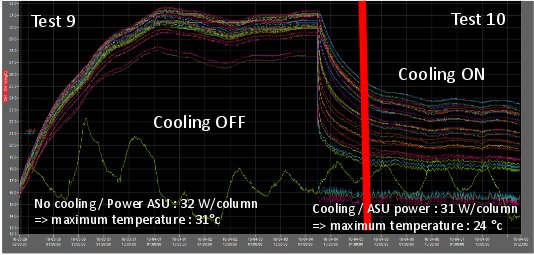}
    \caption{Thermal test results}
    \label{TestResults}
\end{figure}

Figure~\ref{TestResults} shows the first tests results in line 
and was compared with simulations.
These tests demonstrate the important thermal inertia of the system 
to reach the foreseen steady state.
Four days of stabilisation are the minimal time to avoid transient effect.
The cooling is satisfactory with a maximum 
variation of temperature of 2.2$^{\circ}$C for the barrel modules 
and therefore 6$^{\circ}$C for the longest end-cap modules.
These cooling tests validate a correlation with simulations
(transfer coefficients, contacts, conductivities, design of copper foils, 
geometries), and check the
thermal dissipation behavior of the system. 

\subsection{The global cooling system}

\begin{figure}[!h]
    \centering
    \includegraphics[width=15cm]{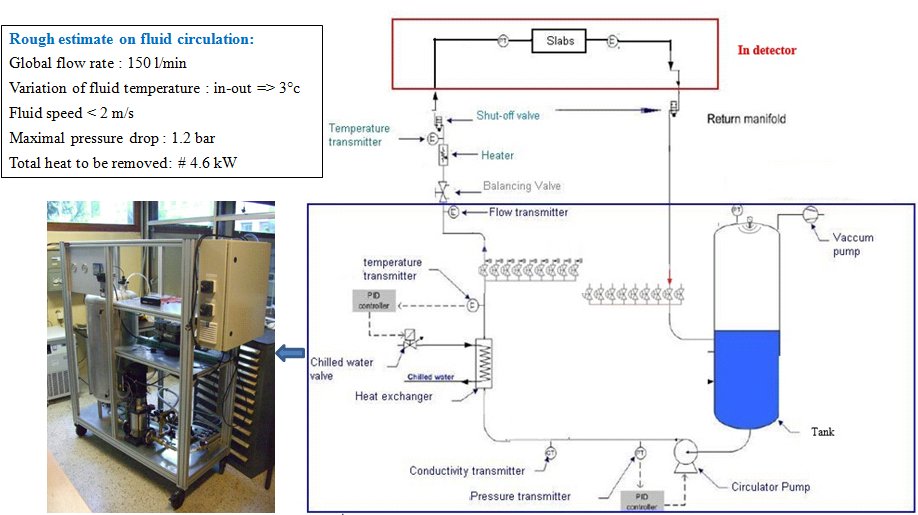}
    \caption{Left: cooling station in constuction. 
           Right: global cooling system.}
    \label{CoolingSystem}
\end{figure}
\begin{figure}[!h]
    \centering
    \includegraphics[width=8cm]{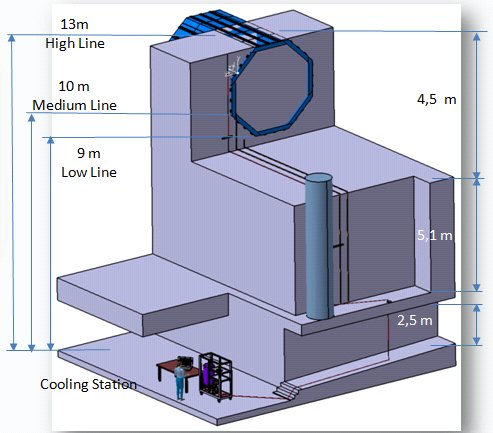}
    \caption{Cooling test area with a drop of 13~m at LPSC.} 
    \label{Loops}
\end{figure}

The detailed study of the global installation  
(circuits, sensors, cooling station, valves, vaccum pumps, etc.) 
and integration of
the pipelines network are underway (see fig.~\ref{CoolingSystem}).
A full scale leak less loops representative 
of the three specific real path types (9~m, 10~m and 13~m), 
through which the detector would be cooled, 
incorporating the head losses due in particular to the crossing of 
the other detectors, will be tested. 
This installation will be implemented at LPSC-Grenoble during spring 2017
(see fig.~\ref{Loops}).

\subsection{Handling tool and integration}

\begin{figure}[!h]
    \begin{minipage}[t]{7cm}
			\centering
			\includegraphics[width=6cm]{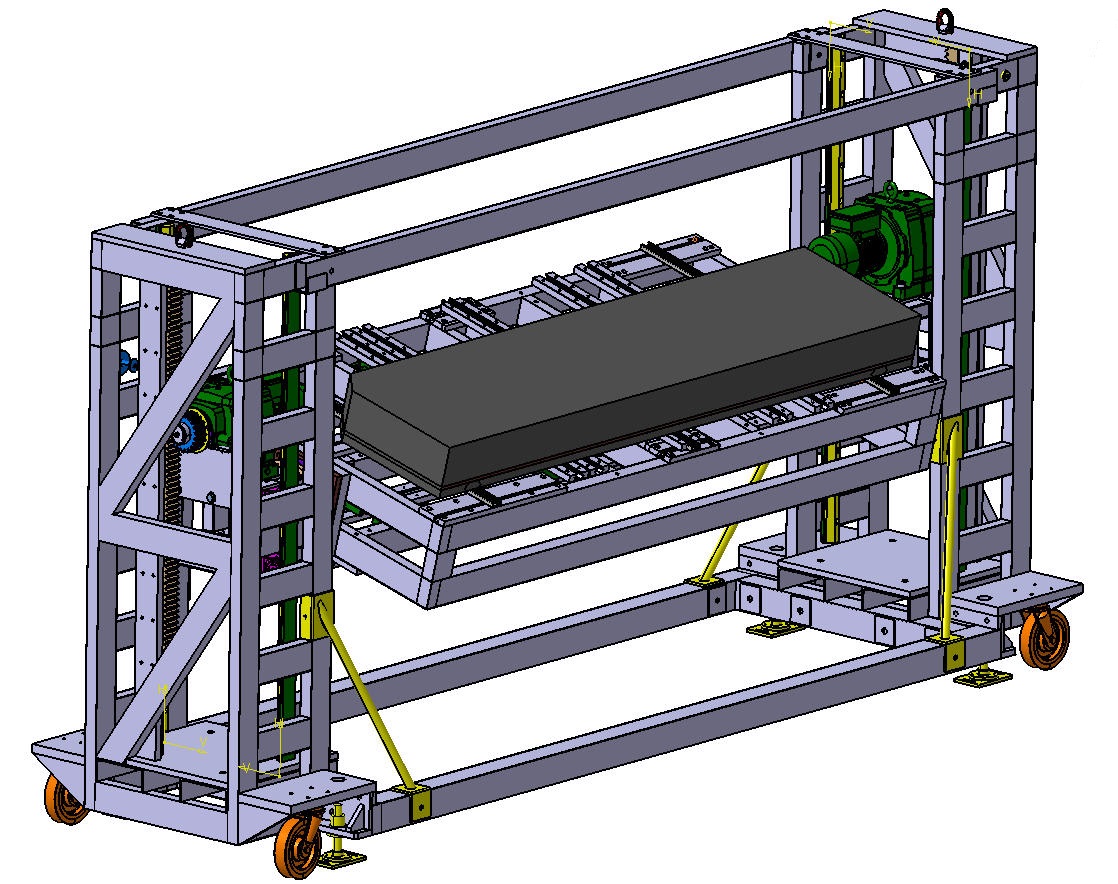}
			\caption{Heavy tool for module integration.}
    \label{HandlingTool}
			\end{minipage}
			\hspace{1mm}
			\begin{minipage}[t]{7cm}
        \centering
				\includegraphics[width=7cm]{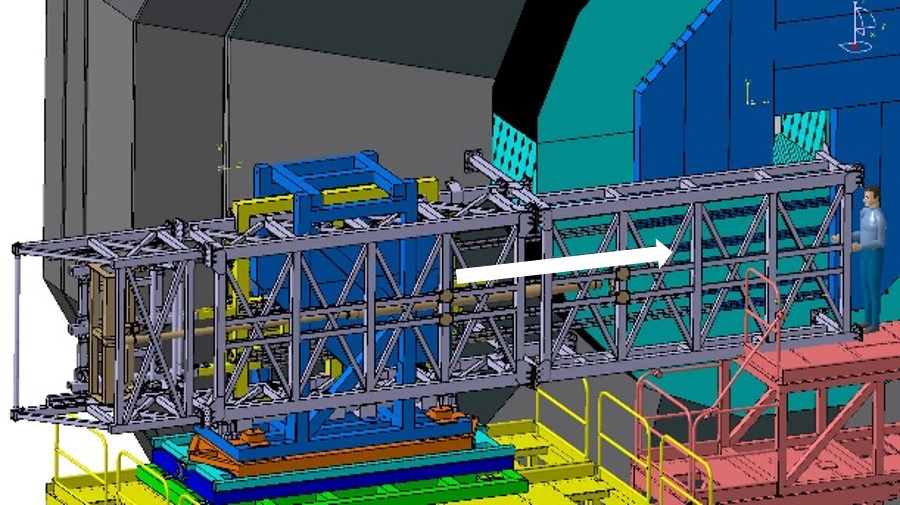}
				\caption{Transfer of ECAL quadrant on quadrant insertion 
           tool (ALICE like).}
    \label{IntegrationTooling}
		  \end{minipage}
\end{figure}

Heavy equipment handling, transport and positioning of modules is 
underway (see fig.~\ref{HandlingTool}).
It will also make possible to test the fastening system
and the integration of the first cooling networks 
on the intrados of the modules.

The different steps of the end-cap integration on site: 
assembly and test of modules and quadrants, quadrant assembly on cradles, 
assembly of quadrants on hadronic end-caps, 
ECal end-cap general cooling integration on HCal,
have been studied in detail, allowing to define the tooling equipment 
(see fig.~\ref{IntegrationTooling}), 
surfaces, time schedule etc., 
needed for each operation. 
\\E.g.:
\\Estimation of surface needed for operations 
\begin{itemize}
\item 500~m$^{2}$ for assembly of modules, 
specific structures, services \& slabs detectors storage
\item 120~m$^{2}$ zone for end-cap insertion on each side 
of whole structure in assembly hall
\end{itemize}
Optimistic time schedule
\begin{itemize}
\item ECAL integration time of one end-cap: 
20 weeks on ILC Campus / 8 weeks in assembly hall
\item Total time for the two end-caps: 56 weeks
\end{itemize}

\section{Conclusion}

This report concludes, to a large extent, the prototyping phase of long CFRP alveolar structure and the design phase of a highly integrated and efficient cooling for the Electromagnetic Calorimeter wich constitutes therefore a milestone within the WP14 AIDA2020 program (cooling topic). Reliable estimation of cooling needs can be obtained by the flexible thermal model of tungsten / carbon fiber calorimeters. The design progression highlights the technological challenge to construct such compact and highly efficient cooling systems embedded in wide composite structures. This concerns in particular the integration of the heat exchanger onto the slabs front end. All elements of the cooling system represent technology reaching beyond nowadays state-of-the-art.
The design allows for the validation of important engineering issues like heat dissipation of the front end electronics embedded into the calorimeter layers, mainly by passive cooling. In parallel to this study, the construction of the large leak less cooling loop will be implemented during spring 2017. The construction of an heavy tool for module integration will allow in 2017 to perform tests on long CFRP structures and on fastening systems too.
Again, the progress in this context will benefit from the collaboration between CALICE and other calorimeter collaborations (ATLAS HGTD, CMS HGCAL).
In conclusion, the CFRP alveolar structure development and the integrated sub-atmospheric water-cooling system will constitute a significant step towards the construction of a Si/W electromagnetic calorimeter for the ILC detector.

\section{Acknowledgements}

The different studies and prototype constructions presented in this 
article have been possible with the help of the financial support of 
Advanced European Infrastructures for Detectors at Accelerators, 
namely AIDA and AIDA-2020 projects.

\begin{figure}[!h]
    \centering
    \includegraphics[width=15cm]{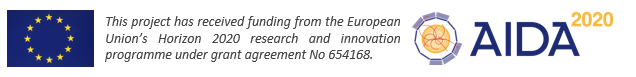}
    \label{AIDA2020-logo}
\end{figure}


\end{document}